\documentstyle[12pt]{article}
\begin{document}
\pagenumbering{arabic}
\setcounter{page}{1}
\newcommand{\bc}{\begin{center}}
\newcommand{\ec}{\end{center}}
\newcommand{\br}{\begin{right}}
\newcommand{\er}{\end{right}}
\bc
{\LARGE \bf Analysis of the anomalous events
       $e^{+}\,e^{-}\,\rightarrow\,l^{+}\, \l^{-}\,\gamma\,\gamma$}
\ec
\vspace{4cm}
\bc
{\bf V.~A.~Litvin} \\
{\em Institute for High Energy Physics \\
 Protvino, Moscow Region 142284, RUSSIA} \\
\vskip0.3cm
and
\vskip0.3cm
{\bf S.~R.~Slabospitsky} \\
{\em Institute for High Energy Physics \\
 Protvino, Moscow Region 142284, RUSSIA} \\
\ec
\newpage
\vskip5.5cm
\begin{abstract}
The $e^{+}\,e^{-}\,\rightarrow\,l^{+}\, \l^{-}\,\gamma\,\gamma$ anomalous
 events, regis\-te\-red at $L3$ de\-tec\-tor at $e^+ e^-$ $CERN-LEP$ collider
have been analysed. It has been shown that the interpretation of such events
as a manifestation of scalar (pseudoscalar) resonance with the mass of 60 GeV
contradicts other experimental data.
\end{abstract}
\newpage
\section{\bf Introduction}

 Recently the $L3$ collaboration ($CERN-LEP$ collider) has reported
four unusual events in the following reaction \cite{1} :
\begin{eqnarray}
e^{+}\,e^{-}\,\rightarrow \,l^{+}\ l^{-}\,\gamma\,\gamma.
\end{eqnarray}
The two photon invariant mass in all these events is about
$60$ GeV.

There are uncertainties in the interpretation of these events. According
to the authors estimates \cite{1} the probability that these events
could be interpreted as the usual QED background is relatively small
($\sim\,10^{-3}$, see \cite{1}).

 The production of the Standard Model Higgs of 60 GeV mass (with subsequent
Higgs decay into two photons) accompanied by $l^+ l^-$ pair could be one
of the possible explanation of the observed effect.
However the probability of such process is very small ($\sim \, 10^{-5}$,
see \cite{2}). Indeed, one should expect the production of about one event of
the Higgs
 boson with $M_H = 60$ GeV at $10^6$ Z--bosons. Taking into account that
$Br(H \to \gamma \gamma)$ is of $10^{-4}$ order of magnitude, the number
of events of reaction (1) for $10^6$ Z--bosons is about $10^{-4}$. This value
is $10^5$ times less than one observed in the $L3$ experiment.

The interpretation of these events via additional Higgs bosons (using
the extension of Standard Model) also meets some difficulties (see \cite{3},
for example).

In this paper the analysis of the reaction (1) is carried out with the
assumption of existence of scalar (pseudoscalar) resonance with mass about
60 GeV. The nature of this resonance  (i.e. whether this
resonance appears to be an elementary object of Higgs boson type or a
composite one of quarkonium type) is not discussed here. The interaction
of such resonance with
photon and Z--boson is considered in general with three arbitrary coupling
constants ($R\gamma\gamma \,\to \, g_{\gamma\gamma};$ $R\gamma Z\,\to \,
g_{\gamma z}$; $RZZ\,\to \,g_{zz}$).
As it follows from this analysis the obtained results do not depend on the
presence or absence of the possible interaction of this resonance with
other particles (quarks, leptons, etc).

A possible production of such resonance in other interactions (for example
in $Z\,\to \, \gamma \, R( \to \gamma \gamma)$ decay) has been analysed
alongside with the process~(1) for the sake of the corresponding coupling
constants estimates. Such combined analysis of different experiments has made
it possible to obtain the upper limits for coupling constants of
resonance interaction with photon and $Z$--boson.

The paper is organized as follows. The resonance interactions with photon and
$Z$--boson are considered in Section~2, and calculations of the necessary
expressions for decay width and production cross section is given.
Different experiments resulting in the appearance of this proposed
resonance are analysed in Section~3. The main obtained results are presented
in Conclusion.

\section{\bf Theoretical estimates}

As it has been mentioned in Introduction the existence of scalar
(pseudoscalar) resonance $R$ with about $60$ GeV mass is assumed for
the analysis of reaction~(1) events.

One can write rather general expressions for the $R\gamma\gamma$,
$R\gamma Z$ and
$RZZ$ vertices taking into account the Lorentz-- and gauge invariance
constraints only~:
\begin{eqnarray}
R^+ \gamma\gamma & = &  \frac{ g_{\gamma\gamma} }{M_{R}}
 (g^{\mu\nu}(k_{1}k_{2})-k_{1}^{\nu}k_{2}^{\mu})
 e_{1}^{\mu}e_{2}^{\nu},    \nonumber\\
R^{-}\gamma\gamma & = & \frac{g_{\gamma\gamma}}{M_{R}}
 \varepsilon^{\mu\nu\alpha\beta}k_{1}^{\mu}k_{2}^{\nu}
 e_{1}^{\alpha}e_{2}^{\beta},  \nonumber\\
R^+ \gamma Z & = &  \frac{ g_{\gamma z} }{M_{R}}
 (g^{\mu\nu}(k_{1}k_{2})-k_{1}^{\nu}k_{2}^{\mu})
 e^{\mu}V^{\nu},    \nonumber\\
R^{-}\gamma Z & = & \frac{g_{\gamma z}}{M_{R}}
 \varepsilon^{\mu\nu\alpha\beta}k_{1}^{\mu}k_{2}^{\nu}
 e^{\alpha}V^{\beta},  \nonumber \\
R^+ ZZ & = &   g_{zz} M_{Z}  g^{\mu\nu} V_1^{\mu}V_2^{\nu}, \nonumber \\
R^{-} ZZ & = & \frac{g_{zz}}{M_{R}}
 \varepsilon^{\mu\nu\alpha\beta}k_{1}^{\mu}k_{2}^{\nu}
 V_{1}^{\alpha}V_{2}^{\beta}, \nonumber
\end{eqnarray}
where $R^{+}(R^{-})$ denotes scalar (pseudoscalar) resonance;
$k_1$ and $k_2$ are momenta of two final photons (photon and $Z$--boson or
two $Z$--bosons); $e^{\nu}(V^{\nu})$ is the photon ($Z$--boson) polarization
vector. The factor $1 / M_R$ takes into account the dimensionless of coupling
constants.

As it was mentioned in Introduction the nature of this resonance is not
discussed. Therefore we do not consider possible types of interactions with
other particles (quarks, leptons, etc.).  Moreover, our analysis does not
depend on the existence of such interactions as it will be shown below.

The basic process (1) cross section can be easily calculated from the mentioned
above expressions for the vertices of resonance interactions with photon and
 $Z$--boson. Four Feynman diagrams describe this process (see Appendix for
details). And cross section of this reaction depends on all the three coupling
constants and also on branching ratio of $R$ resonance decay into
$\gamma\gamma$ :
\begin{eqnarray}
\sigma = f(g_{\gamma\gamma}\,;\,g_{\gamma z}\,;\,g_{zz})Br(R\,
\to\,\gamma\gamma)
\end{eqnarray}
where $Br(\,R\,\rightarrow\,\gamma\,\gamma) \,=
\, \Gamma(\,R\,\rightarrow\,\gamma\,\gamma\,) \, / \, \Gamma_{tot}(R)$.

\noindent The expression for cross section is given in the Appendix.

In order to determine the effective coupling constants values separately we
consider processes which may result in the appearance of hypothetical
$R$--resonance.  Let us also choose those processes in which $R$ can
decay into $\gamma\gamma$.

\subsection {\it Two photon annihilation process}
\begin{eqnarray}
	e^{+}\,e^{-}\,\rightarrow\,e^{+}\,e^{-}\,R
         \,(\,\rightarrow\,\gamma\,\gamma\,)
\end{eqnarray}

Diagrams with $t$--channel photon exchange will contribute substantially
this process far from the $Z$--boson pole.
And $Z$--boson contribution to the $s$--channel will be suppressed for a major
extend. Weak interactions contribution to the rest diagrams is very small
compared to QED contribution. Then we have in the equivalent photon
approximation~:
\[
 \sigma(e^+ e^- \to e^+ e^- R) \, = \,
        {\eta}^2\frac{8{\pi}^2 \,\Gamma(R \to \gamma \gamma)}
         {s \, M_R} \,f(\frac{M_R^2}{s}),
\]
where $\eta \, = \frac{\alpha}{2\pi}\ln(\frac{s}{4m_e^2}); \quad
 f(\omega) \, = \, \frac{1}{\omega}\,((2+\omega)^2\,\ln(\frac{1}
        {\omega})-2(1-\omega)(3+\omega)).$

\noindent The corresponding expression for the process (3) cross section is
the following~:
\begin{eqnarray}
 \sigma(e^{+}\,e^{-}\,\rightarrow\,e^{+}\,e^{-}\,R\,
	(\rightarrow\,\gamma\,\gamma))\,=\,\sigma(e^{+}\,
        e^{-}\,\rightarrow\,e^{+}\,e^{-}\,R) \,Br(R\,\rightarrow\,
	\gamma\,\gamma).
\end{eqnarray}
The decay width of $R$--resonance into two photons is :
\begin{eqnarray}
 \Gamma(R^+\,\rightarrow\,\gamma\,\gamma)\,=\,
 \Gamma(R^-\,\rightarrow\,\gamma\,\gamma)\,=\,
	\,=\, \frac{g_{\gamma\gamma}^{2}}{64\pi}M_{R}.
\end{eqnarray}

\subsection{\it $Z$--boson decay into $3\gamma$}

One can obtain the upper limit for $g_{\gamma z}$ from the experimental
value of $\Gamma(Z\,\rightarrow\,3\gamma)$ \cite{4}~:
\begin{eqnarray}
\Gamma(Z\,\rightarrow\,\gamma\,R(\,\rightarrow\,\gamma\,\gamma))
\,=\,\Gamma(Z\,\rightarrow\,\gamma\,R)\,Br(R\,\rightarrow\,
\gamma\,\gamma)\,\leq\,\Gamma(Z\,\rightarrow\,3\gamma).
\end{eqnarray}
where the expression for $\Gamma(Z\,\rightarrow\,\gamma\,R)$ is as follows~:
\begin{eqnarray}
\Gamma(Z\,\rightarrow\,\gamma\,R^+)\,=\,
\Gamma(Z\,\rightarrow\,\gamma\,R^-)\,=\,
\frac{g_{\gamma z}^2
m_Z\,(1\,-\,\delta)^3}{96\pi\delta},
\end{eqnarray}
here $\delta = (\frac{M_R}{M_Z})^2$.

\subsection {\it $\nu\bar\nu$ pair production with two photons}
\begin{eqnarray}
e^{+}\,e^{-}\,\rightarrow\,\nu\bar\nu\,R\,(\,\rightarrow\,\gamma\,\gamma\,)
\end{eqnarray}

The cross section of this process is calculated analogically to one
of basic process~(1) (see Appendix).

The analogous $\nu\bar\nu\gamma\gamma$ final state can be produced in the
following reaction~:
\[
e^{+}\,e^{-}\,\rightarrow\,\gamma\,R\,(\,\rightarrow\,\gamma\,Z^{\ast}\,
(\,\rightarrow\,\nu\, \bar \nu)).
\]
But the contribution of these diagrams is negligible due to very small
probability of the $R\,\rightarrow\,\gamma\,Z^{\ast}\,
(\,\rightarrow\,\nu\bar\nu\,)$ decay compared to
$R\,\rightarrow\,\gamma\,\gamma\,$ decay.

\section{\bf The analysis of the experimental data}

One can obtain the upper limits for
$g_{\gamma\gamma}$, $g_{\gamma z}$ and $g_{zz}$ coupling constants separately
using the experimental data \cite{1}, \cite{4} and \cite{5}
for the processes (3), (6) and (8).

\subsection {\it $R$--resonance production in $e^+ e^-$ annihilation at
the KEK energies}
Analysing process (3) we used the following experimental information~\cite{5}~:
no events were observed, i.e. $N_{ev}\,\leq\,1\,$;
the total integrated luminosity is ${\cal L}\,=\,60 {\it pb}^{-1}$
and $\sqrt{s}\,=\,70\,$ GeV.

{}From the expression (4) and $N_{ev}\leq 1$ constraint the following
inequality is obtained~:
\[
\ 1.17\cdot{10}^5 g_{\gamma\gamma}^2 Br(R\,\to\,\gamma\,\gamma)
\,\leq\,1
\]
or
\begin{eqnarray}
g_{\gamma\gamma}\sqrt{Br(R\,\to\,\gamma\,\gamma)}\,\leq\,
2.92\cdot{10}^{-3}.
\end{eqnarray}

\subsection {\it $Z\,\to\,3\gamma$ decay}
While analysing this decay we used the following experimental data for
$Z\,\to\,3\gamma$ decay~\cite{4} :
\begin{eqnarray}
 Br(Z\,\to\,3\gamma)\,\leq\,6.6\cdot {10}^{-5}. \nonumber
\end{eqnarray}
We can obtain the following upper limit for coupling constant
$g_{\gamma z}$ from this data and expressions (6) and (7)~:
\begin{eqnarray}
g_{\gamma z}\sqrt{Br(R\to\gamma\gamma)}\,\leq\, 0.036.
\end{eqnarray}

\subsection  {\it $e^{+}e^{-}\,\to\,\nu\bar\nu\gamma\gamma$ reaction}
The upper limits for $g_{zz}$ can be obtained from the value of
the process~(8) cross section on the basis of the following experimental
information~\cite{1}~: no events were observed, i.e. $N_{ev}\,\leq\,1\,$;
the total integrated luminosity is ${\cal L}\,=\,21 {\it pb}^{-1}$ and
$\sqrt{s}\,\simeq\,92\,$ GeV.
As a result we obtain :
\begin{enumerate}
  \item For scalar $R^+$--resonance :
  \[
  (1.64\cdot{10}^{-3}g_{\gamma z}^2+
  4.90 g_{zz}^2)Br(R\,\to\,\gamma\gamma)\,\leq\,1.
  \]
  \item For pseudoscalar $R^-$--resonance :
  \[
  (1.0\cdot{10}^{-3}g_{\gamma z}^2+
  0.47 g_{zz}^2)Br(R\,\to\,\gamma\gamma)\,\leq\,1.
  \]
\end{enumerate}

The $g_{\gamma z}$ coupling constant contribution to the above
inequalities is negligible (see~(10)). It enables to obtain the following
upper limit for the constant~$g_{zz}$~:
\begin{enumerate}
  \item In the case of $R^{+}$ :
  \begin{eqnarray}
  g_{zz}\sqrt{Br(R\,\to\,\gamma\gamma)}\,\leq\, 0.451.
  \end{eqnarray}
  \item In the case of $R^{-}$ :
  \begin{eqnarray}
  g_{zz}\sqrt{Br(R\,\to\,\gamma\gamma)}\,\leq\, 1.46 .
  \end{eqnarray}
\end{enumerate}
\vskip0.5cm

Now we can proceed to the calculation of the number of the reaction~(1)
events with the
help of the expression for the basic process cross section.
We use the following data~\cite{1}, namely :
there are $3$ detected events with ${\mu}^{+}{\mu}^- \gamma \gamma$
 final state at $\sqrt{s}\,\simeq\,92$~GeV and the
total integrated luminosity of ${\cal L} \,=\,27\, {\rm pb}{}^{-1}$
and $(m_{\mu^{+}\mu_{-}})_{min}\,=\,18$~GeV.

 As a result we obtain  the theoretical estimates for the
number of events for basic reaction~(1).
\begin{enumerate}
\item For the scalar $R^+$--resonance :
\begin{eqnarray}
N^+_{th}\,=\,(0.15g_{\gamma\gamma}^2 + 72.8g_{\gamma z}^2 +
0.76g_{zz}^2
-9.16\cdot{10}^{-4}g_{\gamma\gamma}g_{\gamma z} \nonumber\\
+ 9.38\cdot{10}^{-5}g_{\gamma\gamma}g_{zz} + 0.99g_{\gamma z}g_{zz})
Br(R\,\to\,\gamma\gamma).
\end{eqnarray}
\item For the pseudoscalar $R^-$--resonance :
\begin{eqnarray}
N^-_{th}\,=\,(0.10g_{\gamma\gamma}^2 + 49.0g_{\gamma z}^2
+ 0.0684g_{zz}^2 - 5.21\cdot10^{-4}g_{\gamma\gamma}g_{\gamma z} \nonumber\\
- 2.30\cdot{10}^{-5}g_{\gamma\gamma}g_{zz} - 0.247g_{\gamma z}g_{zz})
Br(R\,\to\,\gamma\gamma).
\end{eqnarray}
\end{enumerate}

The theoretical number of events for the process~(1)  can be obtained
substituting of the mentioned limits for coupling constants (9)--(12)
into eqs. (13) and~(14)~. As a result we obtain for the
scalar(pseudoscalar) $R^+(R^-)$--resonance~:
\begin{eqnarray}
N^{+}_{th}(N^{-}_{th})\,\leq\,0.265\,(0.197).
\end{eqnarray}
that gives an order of magnitude less events than measured by the
$L3$--col\-la\-bo\-ra\-tion.

The obtained theoretical estimate eqs. (15) can be
slightly increased choosing $M_{R}\,=\,M_{min}(\gamma \gamma)\,=\,58.2$~GeV
(the minimal measured two photon mass)
instead of $M_R\,=\,60$~GeV.
As a result one can obtain instead of eqs.~(15)
for the scalar(pseudoscalar) $R^+(R^-)$--resonance :
\begin{eqnarray}
N^{+}_{th}(N^{-}_{th})\,\leq\,0.288\,(0.220).
\end{eqnarray}
that is still essentially smaller than experimental value
($N_{exp}\,=\,3$).

It should be noted that the reaction (1) cross section and
the correspondent number of events from~(13), (14) is a logarithmic function
only of the minimal invariant mass of final leptons (see Appendix). Therefore
another choice of $(m_{{\mu}^{+}{\mu}^{-}})_{min}$ (2~GeV, for example)
results in an increase of the cross section (and corresponding number of
events) but no more than two times. That is still smaller than
experimental value.

It is necessary to note that the coupling constants estimates (see eqs.
(9) -- (12)) contain the products of constants squared and branching ratio
of $R$--resonance two photon decay.
The corresponding products of coupling constants and $Br(R\, \to \, \gamma
\gamma)$ enter the final expression (13) and (14) for the number of events.
Thus our analysis is independent of the
 $Br(R\, \to \, \gamma \gamma)$ value and the presence (or absence) of
possible interactions of the studied $R$--resonance with other
particles as well.

\section{\bf Conclusions}

The $e^{+}\,e^{-}\,\rightarrow\,l^{+}\, \l^{-}\,\gamma\,\gamma$ anomalous
 events, regis\-te\-red at $L3$ de\-tec\-tor at $e^+ e^-$ $CERN-LEP$ collider
have been analysed under assumption of general type interactions
of the scalar (pseudoscalar) $R$--resonance with photons and/or $Z$--bosons.

We obtained the upper limits for the effective coupling constants of
interaction of this resonance with photon and $Z$--boson separately
from analysis of the following reactions~:
\begin{enumerate}
\item
$g_{\gamma\gamma}\sqrt{Br(R\,\to\,\gamma\,\gamma)}\,\leq\,
2.92\cdot{10}^{-3}$ from reaction
$e^+\,e^-\,\to\,e^+\,e^-\,\gamma \gamma$ (at KEK energies).
\item $g_{\gamma z}\sqrt{Br(R\to\gamma\gamma)}\,\leq\, 0.036$ from the
process $Z\,\to\, \gamma \gamma \gamma$.
\item
$  g_{zz}\sqrt{Br(R^+(R^-)\to\gamma\gamma)}\leq 0.451(1.46)$
from reaction $e^+\,e^-\,\to\, \nu \bar \nu \gamma \gamma$.
\end{enumerate}

The obtained upper limits of coupling constants make it possible to
estimate the theoretical number of events
($N_{th} = 0.197-0.288$), which an order of magnitude
smaller than the experimental data of $L3$--collaboration ($N_{th}=3$).
Therefore our analysis says that these events might be
only due to usual QED background processes (as it is mentioned in~\cite{1}).
have been analysed. And the interpretation of such events
as a manifestation of scalar (pseudoscalar) resonance with the mass of 60 GeV
contradicts other experimental data.

\vskip0.5cm
{\bf Acknowledgements.}

\noindent We thank Yu.M.Antipov, V.I.Borodulin,  E.M.Levin, A.K.Li\-kho\-ded,
V.F.Obraz\-tsov and A.M.Zaitsev for fruitful
discussion. One of us (S.R.S.) would like also to acknowledge the hospitality
extended to him at Fermilab Theoretical Physics Department, where
this work was completed.

\newpage

\Large
{\bf Appendix}
\normalsize

In this Appendix we present the expressions for the cross-section for
the following reaction :
\[
 e^{+}(l_1)e^{-}(l_2)\,\to\,f(l_3)\bar f(l_4) R(\,\to\,\gamma\gamma),
\]
where the 4--momenta of the particles are given in the parenthesis.

This process is described by four Feynman diagrams. The first (second)
diagram $M_1(M_2)$ corresponds to $e^+ e^-$ annihilation via photon with
production of $R$ and $\gamma^{\ast}(Z^{\ast})\,\to\, f\bar f$.
The third (fourth) one $M_3(M_4)$ corresponds to $e^+ e^-$ annihilation
via $Z$--boson with
production of $R$ and $\gamma^{\ast}(Z^{\ast})\,\to\, f\bar f$.
Below we present the expressions for these amplitudes~:
\begin{enumerate}
\item For scalar resonance production:
\begin{eqnarray*}
M_1 &=& \frac{g_{\gamma\gamma}}{M_R}f^{\alpha\beta}
\frac{1}{q_1^2q_2^2}L_1^{\alpha}L_2^{\beta}; \,\,
M_2 = \frac{g_{\gamma z}}{M_R}f^{\alpha\beta}
\frac{1}{q_1^2 Z_2}d_2^{\beta\nu}L_1^{\alpha}K_2^{\nu}; \\
M_3 &=& \frac{g_{\gamma z}}{M_R}f^{\alpha\beta}
\frac{1}{q_2^2 Z_1}d_1^{\alpha\mu}K_1^{\mu}L_2^{\beta}; \,\,
M_4 = g_{zz}M_Z \frac{1}{Z_1 Z_2} g^{\alpha\beta}
d_1^{\beta\nu}d_2^{\alpha\mu}K_1^{\mu}K_2^{\nu};
\end{eqnarray*}
where the following notations are introduced :
\begin{eqnarray*}
&& f^{\alpha\beta} = (q_1q_2)g^{\alpha\beta}-q_1^{\beta}q_2^{\alpha}, \,
 d_{1(2)}^{\alpha\beta} = g^{\alpha\beta}-\frac{q_{1(2)}^{\alpha}
q_{1(2)}^{\beta}}{M_Z^2}, \\
&& Z_{1(2)} = (q_{1(2)}^2-M_Z^2)+iM_Z\Gamma_Z, \quad
q_1^{\mu} = l_1^{\mu}+l_2^{\mu}, \, q_2^{\mu} = l_3^{\mu}+l_4^{\mu}, \\
&& L_1^{\alpha} = e \bar u(l_1)\gamma^{\alpha}u(-l_2), \,
L_2^{\alpha} = e \bar u(l_3)\gamma^{\alpha}u(-l_4), \\
&& K_1^{\alpha} = \bar u(l_1)\gamma^{\alpha}(a_1+b_1\gamma^5)
u(-l_2), \,
K_2^{\alpha} = \bar u(l_3)\gamma^{\alpha}(a_2+b_2\gamma^5)
u(-l_4), \\
\end{eqnarray*}
\item For pseudoscalar resonance production :
\begin{eqnarray}
M_1 &=& \frac{g_{\gamma\gamma}}{M_R}t^{\alpha\beta}
\frac{1}{q_1^2q_2^2}L_1^{\alpha}L_2^{\beta}; \,\,
M_2 = \frac{g_{\gamma z}}{M_R}t^{\alpha\beta}
\frac{1}{q_1^2 Z_2}d_2^{\beta\nu}L_1^{\alpha}K_2^{\nu}; \nonumber\\
M_3 &=& \frac{g_{\gamma z}}{M_R}t^{\alpha\beta}
\frac{1}{q_2^2 Z_1}d_1^{\alpha\mu}K_1^{\mu}L_2^{\beta}; \,\,
M_4 = \frac{g_{zz}}{M_R} \frac{1}{Z_1 Z_2} t^{\alpha\beta}
d_1^{\beta\nu}d_2^{\alpha\mu}K_1^{\nu}K_2^{\mu}; \nonumber
\end{eqnarray}
\noindent where
$t^{\alpha\beta}\,=\,\varepsilon^{\alpha\beta\lambda\sigma}q_1^{\lambda}
q_2^{\sigma}$, other notations are presented above.
\end{enumerate}

The final expression for the cross section production of scalar
$R^+$--resonance is as follows (we put all fermions massless)~:
\begin{eqnarray*}
\sigma\,=\,\frac{{\alpha}^2}{24\pi\,s}(g_{\gamma\gamma}^2F_1\,+\,
g_{\gamma z}^2 F_2\,+\,g_{zz}^2 F_3\,+\,g_{\gamma\gamma}g_{\gamma z}F_4\,+\,
g_{\gamma\gamma}g_{zz}F_5\,+\,g_{\gamma z}g_{zz}F_6),
\end{eqnarray*}
where
\begin{eqnarray*}
&& F_1\,=\,e_1^2e_2^2A_2; \\
&& F_2\,=\,\frac{g^2e_2^2(a_1^2+b_1^2)}{\varepsilon}
A_2+g^2e_1^2(a_2^2+b_2^2)(A_3+A_4)+2g^2e_1e_2a_1a_2A_4; \\
&& F_3\,=\,\frac{g^4(a_1^2+b_1^2)(a_2^2+b_2^2)}{6\varepsilon}A_1;\quad
F_4\,=\,2ge_1^2e_2a_2A_3; \\
&& F_5\,=\,2g^2e_1e_2a_1a_2A_5;\quad
F_6\,=\,\frac{2g^3e_2a_2(a_1^2+b_1^2)}{\varepsilon}A_6.
\end{eqnarray*}
\begin{eqnarray*}
 A_1\,&=&\,\frac{u_0}{2}(5\delta-23-\lambda)+L_0(2{\delta}^2-14\delta)
+J_0\sqrt{\delta(4-\delta)}; \\
 A_2\,&=&\,\frac{u_0}{6\delta}(3+8\delta-5\lambda-3{\delta}^2+\delta\lambda)
+\frac{L_0}{3\delta}(3{\delta}^2-9\delta-{\delta}^3)-\frac{{u_0}^3}{9\delta} \\
&& + \frac{L_1}{3\delta}(1-{\delta}^2)(1+\delta); \\
 A_3\,&=&\,\frac{u_0}{6\delta}(13\delta-3{\delta}^2+\delta\lambda-6\lambda-7)
+\frac{L_0}{3\delta}(6{\delta}^2-{\delta}^3-18\delta)-\frac{{u_0}^3}{9\delta}
\\
&& -\frac{J_0}{3\delta}(6-4\delta+{\delta}^2)\sqrt{\delta(4-\delta)}; \\
 A_4\,&=&\,\frac{u_0}{6\delta(1-\delta)}(-7\delta^2+24\delta+
\delta\lambda-23)+\frac{L_0}{3\delta}(3\delta^2-12\delta+6) \\
&& +
\frac{J_0}{3\delta}\sqrt{\frac{\delta}{4-\delta}}(3\delta^2-18\delta+30); \\
 A_5\,&=&\,u_0\frac{2\delta-3}{\sqrt{\delta}(4-\delta)}-
\frac{L_0}{\sqrt{\delta}}(2\delta-2)+ J_0\frac{5-\delta}{\sqrt{4-\delta}}; \\
 A_6\,&=&\,\frac{u_0}{2\sqrt{\delta}}(3-3\delta+\lambda)+\frac{L_0}
{\sqrt{\delta}} (4\delta-{\delta}^2)+J_0\sqrt{4-\delta}(2-\delta); \\
 L_0\,&=&\,\ln(\frac{3\sqrt{\delta}}{1+\delta-\lambda-u_0}), \quad
L_1\,=\,\ln(\frac{(1-\delta)^2+(1-\delta)u_0-(1+\delta)\lambda}{2\lambda
\sqrt{\delta}}); \\
 J_0\,&=&\,\frac{\pi}{2}- Asin(\frac{(3-\delta+\lambda)\sqrt{\delta}}{2(1-
\delta)}), \quad
u_0\,=\,\sqrt{(1-\delta)^2-2(1+\delta)\lambda+{\lambda}^2}
\end{eqnarray*}
where
\begin{eqnarray}
&& g\,=\,\frac{1}{\sin(2{\theta}_W)},\,\,\delta\,=\,\frac{M_R^2}{s},\,\,
\varepsilon\,=\,(\frac{\Gamma_Z}{\sqrt{s}})^2,\,\,\lambda\,=\,
(\frac{q_{min}^2}{s}), \nonumber
\end{eqnarray}
 $e_{1(2)}$ is the fermion charge (in electron's charge units);
 $a_{1(2)},b_{1(2)}$ are the vector and axial coupling constants of initial
(final) fermions with $Z$--boson;
$\sqrt{q_{min}^2}\,=\,(m_{{\mu}^{+}{\mu}^{-}})_{min}$ is the minimal value
of the invariant mass of final leptons. The expression for integrated
cross section of pseudoscalar resonance production has the analogous
form as for scalar one, and we do not include it in the paper.
\end{document}